# Three-wave interactions and strange attractor.

S. I. Krasheninnikov and A. R. Knyazev

*University California San Diego, La Jolla, CA, USA*

**Abstract**
It is shown that the incorporation of linear sink/source terms in the three-wave resonance interaction model results in the time dependence of the wave amplitudes, which could exhibit the properties of a strange attractor. This finding demonstrates that the transition to turbulent dynamics of the waves could be related not only to the coupling of wave triads but also to the establishing of the strange attractor-like dynamics within individual wave triads.

Nonlinear interactions of the waves play an important role in different media including liquids, atmosphere, solids, chemical reactions, biological systems, Bose-Einstein condensates, plasma, etc. (e. g. see Ref. 1-13 and the references therein). Such interactions become particularly strong for the case of the resonance wave-wave interactions, which imply some special relations between both frequencies, $\omega_{\vec{k}}$, and the wave-vectors, $\vec{k}$, of interacting waves. For example, the resonance for the case of three-wave ("1", "2", and "3") interactions, which is typical for quadratic nonlinearities of the governing equations, requires $\omega_{\vec{k}_1} = \omega_{\vec{k}_2} + \omega_{\vec{k}_3}$ and $\vec{k}_1 = \vec{k}_2 + \vec{k}_3$.

Three- and four-wave resonance interactions play a crucial role in establishing turbulence spectra in inertial range of the wave-vectors (see Ref. 1 and the references therein) and in the interactions of the waves with desperate scales, see Ref. 11, 12 and the references therein.

We notice that the coupling of different triads of the resonance waves could result in transition to chaotic turbulence regime of wave energy transfer from one wave to another one (see Ref. 13 and the references therein). For example, in Ref. 14 it was suggested that such coupling could trigger violent bursts of so-called Edge Localized Modes in a tokamak.

Most of the studies of the wave-wave interactions assume that the individual waves participating in such interactions are neutrally stable. However, in [15] two-dimensional model of the magnetized plasma flow was considered and it was shown that with the addition of sink/source terms (described by negative/positive growth rates) to the dynamic equation of individual waves could drastically change the dynamics of the whole ensemble of the waves. In particular, it was shown that the variation of the strength of the sink/source terms is causing the transition from a rather quiescent qusisteady state regime to the regime characterized by strong, intermittent fluctuations of plasma energy and enstrophy. Such transition is accompanied by a strong increase of the diffusion coefficient of a passive scalar.

Therefore, to elucidate the role of such sink/source terms in the transition to turbulence here we study their impact on just three-wave resonance interactions.

We consider the amplitude of the fluctuations in the form $\Phi(\vec{r},t) = \sum_{\vec{k}} \left( \Phi_{\vec{k}}(t) \exp(-i\omega_{\vec{k}} t + i\vec{k}\cdot\vec{r}) + c.c. \right)$, we notice that in a general case the amplitudes $\Phi_{\vec{k}}(t)$ could be complex numbers. Assuming the quadratic nonlinearity of the equation governing the amplitudes of the harmonics $\Phi_{\vec{k}}(t)$, for the case of resonance three-wave interactions, $\omega_{\vec{k}} = \omega_{\vec{k}'} + \omega_{\vec{k}-\vec{k}'}$, we have:

$$\frac{d}{dt}\Phi_{\vec{k}}(t) = \sum_{\vec{k}'} M(\vec{k}', \vec{k} - \vec{k}')\Phi_{\vec{k}'}(t)\Phi_{\vec{k}-\vec{k}'}(t) + \Gamma_{\vec{k}}\Phi_{\vec{k}}(t), \tag{1}$$

where $M(\vec{k}', \vec{k} - \vec{k}')$ is the matrix element describing nonlinear interactions of harmonics, whereas $\Gamma_{\vec{k}}$ describe linear growth (for $\Gamma_{\vec{k}} > 0$) or decay (for $\Gamma_{\vec{k}} < 0$) of the modes.

In what follows, we consider the three-wave resonance interaction of the harmonics $\Phi_1(t) \equiv \Phi_{\vec{k}_1}(t)$, $\Phi_2(t) \equiv \Phi_{\vec{k}_2}(t)$, and $\Phi_3(t) \equiv \Phi_{\vec{k}_3}(t)$, described by Eq. (1), assuming that $\vec{k}_1 = \vec{k}_2 + \vec{k}_3$.

In many cases, with a proper normalization of the amplitudes $\Phi_i(t)$ (for simplicity we assume that all amplitudes are the real numbers), we arrive to the following system of equations (e.g. see [1, 8-10, 13]):

$$\begin{cases} \dfrac{da_1}{dt} = a_2 a_3 + \gamma_1 a_1 \\ \dfrac{da_2}{dt} = -a_1 a_3 + \gamma_2 a_2 \\ \dfrac{da_3}{dt} = -a_1 a_2 + \gamma_3 a_3 \end{cases}, \tag{2}$$

where $a_i$ and $\gamma_i$ are the normalized amplitudes and linear growth rates of the harmonics (we notice that $\gamma_i$ are real numbers). We notice that with further renormalization of the amplitudes $a_i$ and time we find that the solutions of Eq. (2) are governed by three parameters. For example, for the renormalization $a_i |\gamma_1| \to a_i$ and $t|\gamma_1| \to t$, these parameters are $\sigma \equiv \gamma_1/|\gamma_1| = \pm 1$, $\hat{\gamma}_2 \equiv \gamma_2/|\gamma_1|$, and $\hat{\gamma}_3 \equiv \gamma_3/|\gamma_1|$. However, keeping this fact in mind, for better tracking of the origin of different terms in our analytic expressions in what follows we will be using Eq. (2) as it is.

For $\gamma_i = 0$ the solutions of Eq. (2) were studied in details (e.g. see Ref. [13] and the references therein) and it is known that for this case Eq. (2) conserves so-called the Manley-Rowe quadratic forms, $A_{12}^2 \equiv a_1^2 + a_2^2$ and $A_{13}^2 \equiv a_1^2 + a_3^2$. However, for $\gamma_i \neq 0$ from Eq. (2) we find

$$\begin{cases} \dfrac{d}{dt} A_{12}^2 = \gamma_1 a_1^2 + \gamma_2 a_2^2 \\ \dfrac{d}{dt} A_{13}^2 = \gamma_1 a_1^2 + \gamma_3 a_3^2 \end{cases}. \tag{3}$$

From Eq. (3) it follows that in order to avoid monotonously increasing (decreasing) values of $A_{12}^2$ and $A_{13}^2$, we should have either

$$\gamma_1 > 0 \text{ and } \gamma_2, \gamma_3 < 0, \tag{4}$$

or

$$\gamma_1 < 0 \text{ and } \gamma_2, \gamma_3 > 0. \tag{5}$$

However, even though inequalities (4) or (5) are fulfilled the values $A_{12}^2$ and $A_{13}^2$ still could increase monotonically since Eq. (2) allows particular exponentially growing time-dependent solutions: ($\hat{a}_1(t) \propto e^{\gamma_1 t}, 0, 0$), for the case (4), and ($0, \hat{a}_2(t) \propto e^{\gamma_2 t}, 0$) and ($0, 0, \hat{a}_3(t) \propto e^{\gamma_3 t}$), for the case (5). But, an unlimited growth of $A_{12}^2$ and $A_{13}^2$, associated with exponentially growing solutions, could be avoided providing that the solutions $\hat{a}_i(t)$ are unstable. We consider the stability of the solutions $\hat{a}_i(t)$ by keeping only linear terms in Eq. (2) with respect to small perturbations: $\tilde{a}_2(t) \propto \tilde{a}_3(t) \propto \exp[S_{23}(t)]$ for the solution ($\hat{a}_1(t), 0, 0$); $\tilde{a}_1(t) \propto \tilde{a}_3(t) \propto \exp[S_{13}(t)]$ for the solution ($0, \hat{a}_2(t), 0$); and $\tilde{a}_1(t) \propto \tilde{a}_2(t) \propto \exp[S_{12}(t)]$ for the solution ($0, 0, \hat{a}_3(t)$). As a result, we find

$$\dot{S}_{23} = \pm \hat{a}_1(t), \tag{6}$$

and

$$\dot{S}_{13} \approx \frac{\gamma_1 + \gamma_3}{2} \pm i\hat{a}_2(t), \qquad \dot{S}_{12} \approx \frac{\gamma_1 + \gamma_2}{2} \pm i\hat{a}_3(t). \tag{7}$$

From (6) we see that for $\dot{S}_{23} > 0$ the solution ($\hat{a}_1(t), 0, 0$) is unstable and the instability growth rate is faster than the growth rate of the function $\hat{a}_1(t)$. Whereas from (7), recalling Eq. (5), we see that the fastest growth rate of the instability of the solutions ($0, \hat{a}_2(t), 0$) and ($0, 0, \hat{a}_3(t)$) is smaller than fastest growth rate of the time dependent components $\hat{a}_2(t)$ and $\hat{a}_3(t)$. As a result, we conclude that for the case (4) unlimited growth of $A_{12}^2$ and $A_{13}^2$, caused by the solution ($\hat{a}_1(t) \propto e^{\gamma_1 t}, 0, 0$) could be interrupted by an onset of the instability of such solution. Whereas for the case (5) the instabilities of the solutions ($0, \hat{a}_2(t), 0$) and ($0, 0, \hat{a}_3(t)$) are too slow to prevent an unlimited growth of $A_{12}^2$ and $A_{13}^2$.

Therefore, in what follows we only consider the values $\gamma_i$ satisfying inequalities (4). For this case, we find that apart from the time dependent solution ($\hat{a}_1(t) \propto e^{\gamma_1 t}, 0, 0$), Eq. (2) has some stationary solutions, $a_i \equiv \bar{a}_i$. After simple algebra we find that in addition to a trivial solution $\bar{a}_i = 0$, there are other stationary solutions determined by:

$$\bar{a}_1^2 = \gamma_2 \gamma_3, \qquad \bar{a}_2^2 = -\gamma_1 \gamma_3, \qquad \bar{a}_3^2 = -\gamma_1 \gamma_2. \tag{8}$$

However, in addition to Eq. (8) such solutions should satisfy the relation

$$\bar{a}_1 \bar{a}_2 \bar{a}_3 = -\gamma_1 \gamma_2 \gamma_3 \equiv -\gamma_g^3 < 0. \tag{9}$$

Whereas Eq. (8) gives the magnitudes of $\bar{a}_i$, Eq. (9) selects the signs of $\bar{a}_i$, which satisfy stationary solutions. As a result, taking into account Eq. (8, 9) we find that for real $a_i$ Eq. (2) has one trivial, $\bar{a}_i = 0$, and four other stationary solutions defined by Eq. (8, 9).

Next, we consider the stability of these stationary solutions. From Eq. (2, 4) it is obvious that the trivial solution $\bar{a}_i = 0$ is unstable since at least one of $\gamma_i$ is positive. To consider the stability of stationary solution defined by Eq. (8, 9) we take $a_i = \bar{a}_i + \tilde{a}_i \exp(\omega t)$, where $\tilde{a}_i$ are the small perturbations. Then, from the linearized version of Eq. (2), we come to the following equation for $\omega$:

$$D \equiv \begin{Vmatrix} \omega - \gamma_1 & -\bar{a}_3 & -\bar{a}_2 \\ \bar{a}_3 & \omega - \gamma_2 & \bar{a}_1 \\ \bar{a}_2 & \bar{a}_1 & \omega - \gamma_3 \end{Vmatrix} = 0. \tag{10}$$

Taking into account Eq. (7, 9) from Eq. (10) we find

$$F(\omega) \equiv \omega^2(\omega - 3\gamma_a) + 4\gamma_g^3 = 0, \tag{11}$$

where $\gamma_a = (\gamma_1 + \gamma_2 + \gamma_3)/3$. The discriminant of Eq. (11) is

$$\Delta = 432\gamma_g^3(\gamma_a^3 - \gamma_g^3). \tag{12}$$

We recall that for $\Delta > 0$ ($\Delta < 0$) cubic equation (11) has three real roots (one real and two complex conjugates). Taking into account Eq. (9), the dependence the function $F(\omega)$ on $\omega$, and the expression (12), we find that: (i) for $\gamma_a^3 > \gamma_g^3 > 0$, which corresponds to $\Delta > 0$, Eq. (11) has two positive and one negative real roots; (ii) for $\gamma_g^3 > \gamma_a^3 > 0$, which corresponds to $\Delta < 0$, Eq. (11) has one real negative and two complex conjugate roots; (iii) for $\gamma_g^3 > 0$ and $\gamma_a^3 < 0$, which corresponds to $\Delta < 0$, Eq. (11) has one real negative and two complex conjugate roots;.

The cases (ii) and (iii) have one real negative and two complex conjugate roots, therefore, we should consider the stability of these cases more thoroughly. We start with the case (ii). From Eq. (11) we find that for $\gamma_g^3 \gtrsim \gamma_a^3 > 0$

$$\omega \approx 2\gamma_a \pm i\sqrt{3\gamma_a(\gamma_g - \gamma_a)}, \tag{13}$$

whereas for $\gamma_g^3 >> \gamma_a^3 > 0$,

$$\omega \approx \frac{1 \pm i\sqrt{3}}{2^{1/3}}\gamma_g, \tag{14}$$

which show that the case (ii) is unstable. Next we consider the case (iii). From Eq. (11) we find that for $\gamma_g^3 >> |\gamma_a|^3$ the solutions having positive real part of $\omega$ is given by Eq. (14). In opposite case, $0 < \gamma_g^3 << |\gamma_a|^3$, in addition to an obvious solution with negative real part, $\omega \approx 3\gamma_a < 0$, Eq. (11) has two small, $|\omega| << |\gamma_a|$, complex conjugate roots with a positive real part:

$$\omega \approx \gamma_g \left(\frac{\gamma_g}{|\gamma_a|}\right)^{1/2} \left\{ \frac{2}{9}\left(\frac{\gamma_g}{|\gamma_a|}\right)^{3/2} \pm i\frac{2}{\sqrt{3}} \right\}, \tag{15}$$

and relatively large imaginary one

$$\left|\frac{\text{Re}(\omega)}{\text{Im}(\omega)}\right| \sim \left(\frac{\gamma_g^3}{|\gamma_a|^3}\right)^{1/2} << 1. \tag{16}$$

Thus, from the cases (i)-(iii) and Eq. (13-15), we see that all stationary solutions defined by Eq. (8, 9) are unstable for any relations between $\gamma_g > 0$ and $\gamma_a$.

The nonlinear solutions of Eq. (2) for different relations between $\gamma_g > 0$ and $\gamma_a$ we obtain numerically. For the case $\gamma_g^3 > 0$ and $\gamma_a^3 < 0$ these solutions, originated in some vicinity of stationary points (8, 9), exhibit the properties of a strange attractor (e.g. see Refs. 16-19) for a wide range of $\gamma_i$. In Fig. 1 one can see the time variation of the amplitudes $a_i(t)$, found from numerical solution of Eq. (2), for $\gamma_1 = 0.01$, $\gamma_2 = -1$, $\gamma_3 = -1.5$.

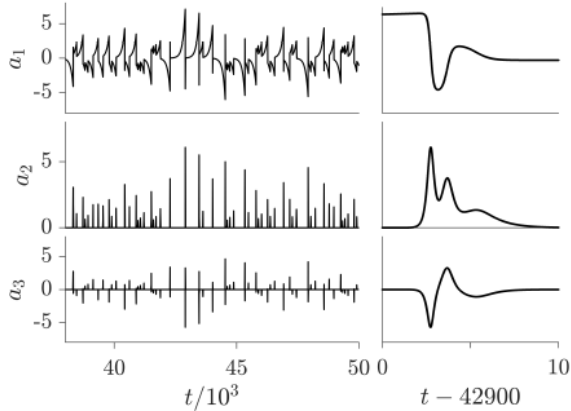
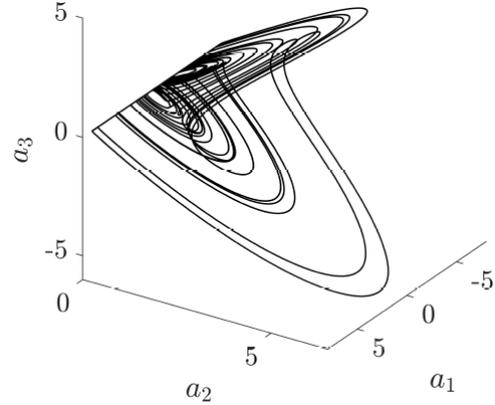

Fig. 1. The variation of $a_i(t)$ found from the numerical solution of Eq. (2) for $\gamma_1 = 0.01$, $\gamma_2 = -1$, $\gamma_3 = -1.5$.

Fig. 2. The phase portrait of $a_i(t)$ found from the numerical solution of Eq. (2) for $\gamma_1 = 0.01$, $\gamma_2 = -1$, $\gamma_3 = -1.5$.

The phase portrait of this solution, time variation of the Manley-Rowe quadratic forms and their difference are shown, correspondingly, in Fig. 2 and Fig. 3. The phase portrait of the solution of Eq. (2) for $\gamma_1 = 0.3$, $\gamma_2 = -1$, $\gamma_3 = -1.5$ is shown in Fig. 4.

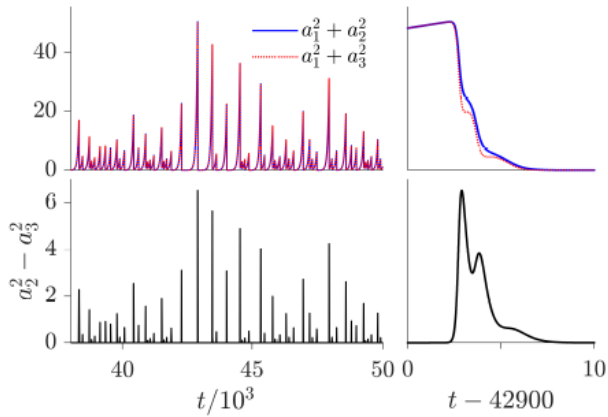
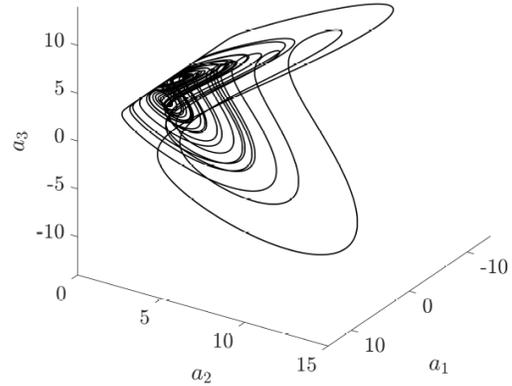

Fig. 3. The time variation of the Manley-Rowe quadratic forms and their difference, found from the numerical solution of Eq. (2) for $\gamma_1 = 0.01$, $\gamma_2 = -1$, $\gamma_3 = -1.5$.

Fig. 4. The phase portrait of $a_i(t)$ found from the numerical solution of Eq. (2) for $\gamma_1 = 0.3$, $\gamma_2 = -1$, $\gamma_3 = -1.5$.

The Lorenz map of the amplitude $a_2(t)$ ($\max_{n+1}(a_2)$ vs. $\max_n(a_2)$), obtained from numerical solutions of Eq. (2) for $\gamma_1 = 0.01$, $\gamma_2 = -1$, $\gamma_3 = -1.5$ and $\gamma_1 = 0.3$, $\gamma_2 = -1$, $\gamma_3 = -1.5$ are shown, correspondingly, in Fig. 5 and Fig. 6. We notice that unlike other strange attractors (e.g. see Ref. 16), only some part of the Lorenz maps shown in Fig. 5, 6, exhibits the "tent map" [19] property. Whereas a straight line, which comes from consecutive maxima of decreasing magnitudes shown in a "blown-up" part of $a_2(t)$ in Fig. 1, gives the rest of the Lorenz maps.

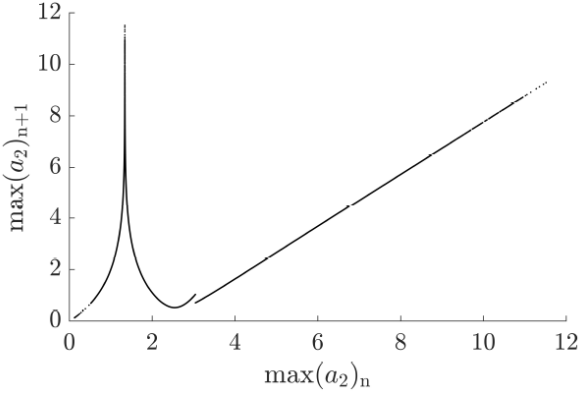 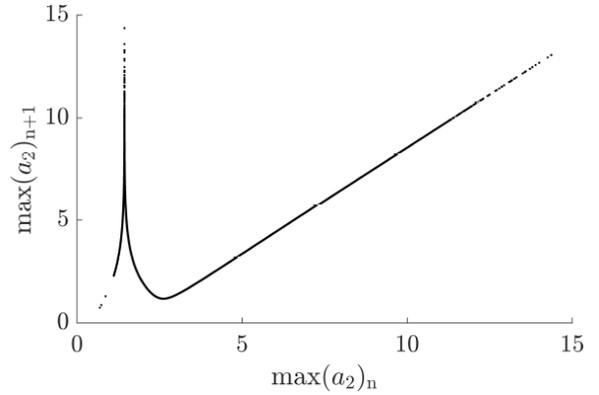

Fig. 5. The Lorenz map of the amplitude $a_2(t)$ ($\max_{n+1}(a_2)$ vs. $\max_n(a_2)$), obtained from the numerical solutions of Eq. (2) for $\gamma_1 = 0.01$, $\gamma_2 = -1$, $\gamma_3 = -1.5$.

Fig. 6. The Lorenz map of the amplitude $a_2(t)$ ($\max_{n+1}(a_2)$ vs. $\max_n(a_2)$), obtained from the numerical solutions of Eq. (2) for $\gamma_1 = 0.3$, $\gamma_2 = -1$, $\gamma_3 = -1.5$.

For a higher value of $\gamma_1$, but keeping $\gamma_2 = -1$, $\gamma_3 = -1.5$, our numerical simulations show that the trajectories $a_i(t)$ originated in the vicinity of the stationary points (8, 9) are approaching the limit cycle (e.g. see Fig. 7 for the case $\gamma_1 = 0.5$, $\gamma_2 = -1$, $\gamma_3 = -1.5$). Further increase of $\gamma_1$ results in unbounded frequently oscillating solutions for $a_i(t)$ (e.g. see Fig. 8 for the case $\gamma_1 = 1$, $\gamma_2 = -1$, $\gamma_3 = -1.5$). However, we should recall that the real governing parameters of Eq. (2) are $\sigma \equiv \gamma_1/|\gamma_1| = \pm 1$, $\hat{\gamma}_2 \equiv \gamma_2/|\gamma_1|$, and $\hat{\gamma}_3 \equiv \gamma_3/|\gamma_1|$.

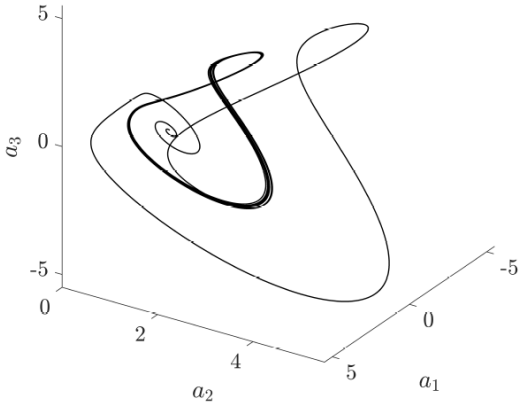 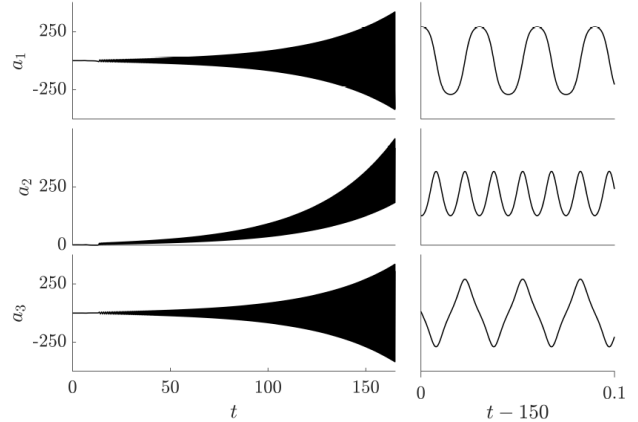

Fig. 7. The phase portrait of $a_i(t)$ approaching the limit cycle found from the numerical solution of Eq. (2) for $\gamma_1 = 0.5$, $\gamma_2 = -1$, $\gamma_3 = -1.5$.

Fig. 8. Time dependence of $a_i(t)$ found from the numerical solution of Eq. (2) for $\gamma_1 = 1$, $\gamma_2 = -1$, $\gamma_3 = -1.5$

In summary, we find that the incorporation of linear sink/source terms into the three-wave resonance interaction model described by Eq. (2), for $\gamma_1 > 0$ and $\gamma_2, \gamma_3 < 0$ results in time-dependent evolution of the normalized wave amplitudes $a_i(t)$, since all stationary solutions of Eq. (2) appear to be unstable. For relatively small $\gamma_1$ ($\gamma_1 \lesssim 0.3$, $\gamma_2 = -1$, $\gamma_3 = -1.5$) these solutions exhibit the properties of a strange attractor. For larger $\gamma_1$ ($\gamma_1 \sim 0.5$, $\gamma_2 = -1$, $\gamma_3 = -1.5$) the solutions are approaching limit cycle, whereas for even larger $\gamma_1$ ($\gamma_1 \gtrsim 1$, $\gamma_2 = -1$, $\gamma_3 = -1.5$) $a_i(t)$ become unbounded oscillating functions. This finding demonstrates that the transition to turbulent dynamics of the waves could be related not only to the coupling of wave triads (e.g. see Ref. 13) but also to the establishing of the strange attractor-like dynamics within individual wave triads.

**Acknowledgments.** This work was supported by the U.S. Department of Energy, Office of Science, Office of Fusion Energy Sciences under Award No. DE-FG02-04ER54739 at UCSD.

**Data availability.** The data that support the findings of this study are available from the corresponding author upon reasonable request.

# References

[1] V. E. Zakharov, V. S. L'vov, and G. Falkovich. "Kolmogorov Spectra of Turbulence", (Series in Nonlinear Dynamics, Springer-Verlag, New York, 1992).

[2] D. C. Fritts and M. J. Alexander, "Gravity Wave Dynamics and Effects in the Middle Atmosphere", Rev. Geophys., **41** (2003) 1-64.

[3] C. P. Connaughton, B. T. Nadiga, S. V. Nazarenko and B. E. Quinn, "Modulational instability of Rossby and drift waves and generation of zonal jets", J. Fluid Mech. **654** (2010) 207–231.

[4] V. I. Erofeev, "Wave processes in solids with microstructure", (Singapore: World Scientific, 2003).

[5] Y. Kuramoto, "Chemical Oscillations, Waves and Turbulence", (Springer, New York, 1984).

[6] N. B. Janson, "Nonlinear dynamics of biological systems", Contemp. Phys. 53 (2012) 137-168.

[7] R. Carretero-Gonzalez, D. J. Frantzeskakis and P. G. Kevrekidis, "Nonlinear waves in Bose–Einstein condensates: physical relevance and mathematical techniques", Nonlinearity **21** (2008) R139–R202.

[8] B. B. Kadomtsev. "Plasma Turbulence" (Acad. Press, London, 1965).

[9] R. Z. Sagdeev and A. A. Galeev. "Nonlinear Plasma Theory", (Benjamin, New York, 1969).

[10] V. N. Tsytovich, "Nonlinear Effects in Plasma", (Plenum, New York, 1970).

[11] A. A. Vedenov, A. V. Gordeev, and L. I. Rudakov, "Oscillations and instability of a weakly turbulent plasma", Plasma Phys. **9** (1967) 719-735.

[12] P. H. Diamond, S.-I. Itoh, K. Itoh, & T. S. Hahm, "Zonal flows in plasma: a review", Plasma Phys. Control. Fusion **47** (2005), R35–R161.

[13] E. Kartashova, "Nonlinear Resonance Analysis, Theory, Computation, Application", (Cambridge University Press, 2010).

[14] J. Dominski and A. Diallo, "Transition of a network of nonlinear interactions into a regime of strong nonlinear fluctuations: A paradigm for the edge localized mode onset", Phys. Plasmas **28** (2021) 092306.

[15] S. A. Galkin, and S. I. Krasheninnikov, "On the sink-source effects in two-dimensional plasma turbulence", Phys. Plasmas **8** (2001) 5091-5095.

[16] E. D. Lorenz, "Deterministic nonperiodic flow", J. Atmosph. Sci. **20** (1963) 130-141.

[17] O. E. Rössler, "An equation for hyperchaos", Phys. Lett. **71A** (1979) 155-157.

[18] G. A. Leonov, N. V. Kuznetsov, "On differences and similarities in the analysis of Lorenz, Chen, and Lu systems", Appl. Math. Comp. **256** (2015) 334-343.

[19] H.-O. Peitgen, H. Jürgens, D. Saupe, "Chaos and Fractals: New Frontiers of Science", Springer, 2004.